\documentclass[aps,prd,reprint,balancelastpage,floatfix,nofootinbib,amsmath,amssymb,superscriptaddress,longbibliography]{revtex4-1}
\usepackage{xcolor}
\usepackage{graphicx}
\usepackage[nopatch=footnote]{microtype}
\usepackage{hyphenat}
\allowdisplaybreaks
\usepackage{bm}         
\usepackage[colorlinks,linkcolor=blue]{hyperref}
\newcommand{\nc}{\newcommand*}
\nc{\Eq}[1]{Eq.~\eqref{#1}}     
\nc{\Fig}[1]{Fig.~\ref{#1}}     
\nc{\Table}[1]{Table~\ref{#1}}  
\nc{\Sec}[1]{Sec.~\ref{#1}}     
\nc{\red}[1]{\textcolor{red}{#1}}
\nc{\dt}{\delta}
\nc{\bt}{\mathbf{t}}
\nc{\eg}{\textit{e.g.~}}
\nc{\bn}[1]{\dt\bm{t}_{\text{#1}}}
\nc{\be}{\bm{\epsilon}}
\nc{\yr}{\mathrm{yr}}

\def\e{
\begin{equation}}
  \def\q{
\end{equation}}

\begin{document}

\title{Primordial Black Holes from Vector-Induced Curvature Perturbations Sourced by Primordial Magnetic Fields}

\author{Chang Han}
\affiliation{Department of Physics,  Key Laboratory of Low Dimensional Quantum Structures
and Quantum Control of Ministry of Education, and Hunan Research Center of the Basic Discipline for Quantum Effects and Quantum Technologies, Hunan Normal University, Changsha, Hunan 410081, China}
\author{Zu-Cheng Chen}
\affiliation{Department of Physics,  Key Laboratory of Low Dimensional Quantum Structures
and Quantum Control of Ministry of Education, and Hunan Research Center of the Basic Discipline for Quantum Effects and Quantum Technologies, Hunan Normal University, Changsha, Hunan 410081, China}

\author{Hongwei Yu}
\email{hwyu@hunnu.edu.cn}
\affiliation{Department of Physics,  Key Laboratory of Low Dimensional Quantum Structures
and Quantum Control of Ministry of Education, and Hunan Research Center of the Basic Discipline for Quantum Effects and Quantum Technologies, Hunan Normal University, Changsha, Hunan 410081, China}

\author{Puxun Wu}
\email{pxwu@hunnu.edu.cn}
\affiliation{Department of Physics,  Key Laboratory of Low Dimensional Quantum Structures
and Quantum Control of Ministry of Education, and Hunan Research Center of the Basic Discipline for Quantum Effects and Quantum Technologies, Hunan Normal University, Changsha, Hunan 410081, China}

\begin{abstract}

Generating an appreciable abundance of primordial black holes (PBHs) requires a substantial enhancement of primordial curvature perturbations on small scales. In this Letter, we propose a new post-inflationary mechanism in which such an enhancement arises during a stiff, or kination, epoch. The mechanism is driven by metric vector perturbations sourced by the vector component of the electromagnetic stress-energy tensor associated with primordial magnetic fields (PMFs). Since these first-order vector modes remain approximately constant during kination, they act as persistent nonlinear sources for second-order scalar perturbations. We show that the resulting vector-induced curvature perturbations are amplified toward the infrared cutoff of the kination band and exhibit the characteristic scaling $\mathcal P_{\mathcal R}(k)\propto k^{-5}$. As a concrete realization, we consider PMFs generated in a Ratra-type magnetogenesis scenario and find that the induced curvature perturbations can produce PBHs with an abundance large enough to constitute a substantial fraction of the dark matter.

\end{abstract}

\maketitle
\textit{Introduction.} Primordial black holes (PBHs), formed through the gravitational collapse of large primordial density fluctuations in the early Universe~\cite{Hawking:1971ei,Carr:1974nx}, provide a unique probe of small-scale physics~\cite{Bird:2016dcv,Sasaki:2016jop,Chen:2021nxo,Liu:2022bvr,Bian:2025ifp,Stojkovic:2004hz,Stojkovic:2005zh,Dong:2015yjs,Khlopov:2008qy,Belotsky:2014kca} and may constitute  part or all of the present dark matter abundance~\cite{Carr:2020gox,Barnacka:2012bm,Montero-Camacho:2019jte,Katz:2018zrn,Niikura:2017zjd}. 
Their formation typically requires a  significant enhancement of primordial curvature perturbations on scales much smaller than those probed by the cosmic microwave background (CMB), while the perturbations on CMB scales must remain consistent with observational constraints~\cite{Sasaki:2018dmp,Planck:2018vyg}.

A variety of mechanisms have been proposed to amplify small-scale curvature perturbations during inflation, including  a transient reduction  of the inflaton rolling speed~\cite{Kinney:1997ne,Inoue:2001zt,Kinney:2005vj,Fu:2019ttf,Fu:2020lob}, variations of the sound speed~\cite{Kamenshchik:2018sig,Ballesteros:2018wlw,Zhai:2022mpi,Zhai:2023azx}, resonant effects~\cite{Cai:2018tuh,Chen:2019zza,Zhou:2020kkf,Chen:2024gqn}, null-energy-condition violation~\cite{Cai:2023uhc} and  isocurvature fluctuations~\cite{Palma:2020ejf,Fumagalli:2020adf}.  PBH formation can also be triggered after inflation. For example, when the inflaton oscillates around the minimum of its potential, resonant amplification of fluctuations in light fields coupled to the inflaton may generate sufficiently large small-scale density perturbations~\cite{Kofman:1994rk,Kofman:1997yn,Green:2000he}. Early-Universe phase transitions provide another route to producing overdense regions that may collapse into PBHs~\cite{Lewicki:2024ghw,Kodama:1982sf,Liu:2022lvz}.
 
 Primordial magnetic fields (PMFs) are another well-motivated relic of the early Universe. They can be generated during phase transitions~\cite{Hogan:1983zz,Vachaspati:1991nm} or during inflation, for instance through mechanisms that break the conformal invariance of electromagnetism~\cite{Kobayashi:2019uqs,Ratra:1991bn,Martin:2007ue,Subramanian:2015lua,Yokoyama:2015era,Vachaspati:1991nm}. PMFs are also of considerable phenomenological interest because they may seed the magnetic fields observed on galactic and cluster scales~\cite{Widrow:2002ud,Vallee:2004osq}.

Observations of TeV blazars by Fermi have been interpreted as providing lower bounds on intergalactic magnetic fields, of order $\gtrsim 10^{-7}\;{\rm nG}$~\cite{Neronov:2010gir}.  
At the same time, PMFs can leave imprints on CMB temperature and polarization anisotropies, stochastic gravitational-wave backgrounds, magnetic reheating, and big-bang nucleosynthesis (BBN)~\cite{Saga:2018qrx,Saga:2017vnm,Kawasaki:2012va,Jedamzik:1999bm,Planck:2015zrl,Maiti:2026magnetism}.
 For example, Planck CMB data constrain the PMF amplitude to be below a few nanogauss at the scale of $1\;{\rm Mpc}$~\cite{Planck:2015zrl}, while CMB spectral distortions constrain magnetic fields on smaller scales, roughly $400\,{\rm pc}\sim 0.6\, {\rm Mpc}$, to be below about $30\,{\rm nG}$~\cite{Jedamzik:1999bm}. Detailed BBN calculations give an upper bound on the present PMF amplitude of approximately $1.5\times10^3\;{\rm nG}$~\cite{Kawasaki:2012va}. More recently, pulsar-timing-array data, including NANOGrav 15-year, EPTA DR2full, and PPTA DR3, have been used to constrain PMFs generated in the early Universe, with characteristic strengths of order $\mathcal{O}(1)\,\mu{\rm G}$ on scales of order $\mathcal{O}(1)\,{\rm pc}$~\cite{Li:2023yaj}. These results suggest that PMFs, especially on small scales, may play an important role in early-Universe cosmology.

 The electromagnetic stress-energy tensor associated with PMFs is quadratic in the magnetic field and can be decomposed geometrically into scalar, vector, and tensor components according to their transformation properties on constant-time hypersurfaces~\cite{Bardeen:1980kt,Kodama:1985bj,Mack:2001gc,Paoletti:2008ck}. 
These components source scalar perturbations, vector perturbations, and gravitational waves, respectively. Previous studies have shown that density perturbations sourced directly by strong magnetic fields may become sufficiently large to form PBHs~\cite{Saga:2020ics,Kushwaha:2024zhd,Maiti:2025pbh,Maiti:2026ultralight}. In this Letter, we point out a different channel: the vector component of the PMF electromagnetic stress-energy tensor sources metric vector perturbations, and these vector modes can in turn induce scalar curvature perturbations at second order.
 
 In standard radiation- or matter-dominated eras, vector metric perturbations decay with the cosmic expansion~\cite{Durrer:1999bk,Paoletti:2008ck}. The situation changes, however, if the Universe undergoes a stiff post-inflationary epoch. Such an epoch can arise when the energy density of the inflaton is dominated by its kinetic energy, leading to a kination phase. Kination appears naturally in quintessential inflation~\cite{Peebles:1998qn,Bettoni:2021qfs} and in a broad class of non-oscillatory inflationary scenarios~\cite{Felder:1999pv,Ellis:2020krl}. The impact of kination on vector-induced gravitational waves has recently been investigated in Ref.~\cite{Bhaumik:2025kuj}. Motivated by these developments, we examine whether magnetically sourced vector perturbations during kination can provide a new origin for the large scalar perturbations required for PBH formation.

In this Letter, we demonstrate that metric vector perturbations sourced by the vector component of the PMF electromagnetic stress-energy tensor can induce enhanced second-order curvature perturbations during a post-inflationary kination era. These induced  curvature perturbations  create over-dense regions that may collapse gravitationally into black holes.  This provides a new PBH production mechanism in which the relevant enhancement occurs not in the linear scalar sector, as in most conventional scenarios, but through second-order scalar perturbations induced by magnetically sourced vector modes. 
We further show that the resulting curvature power spectrum is enhanced toward the infrared cutoff and exhibits a characteristic power-law behavior with spectral index exactly $-5$. Applying the mechanism to PMFs generated in a Ratra-type magnetogenesis scenario, we find that the resulting PBHs can account for a substantial fraction of the dark matter.

\textit{First-order vector perturbations.} 
We first  introduce the first-order vector mode sourced by the vector component of the electromagnetic stress-energy tensor (ESET).   In a spatially flat FLRW background,  vector perturbations $V_i$ can be expressed in the flat gauge  as
\begin{equation}
  ds^2=a^2\left(-d\tau^2+2V_i d\tau dx^i+\delta_{ij}dx^i dx^j\right)
\end{equation}
 where  $\tau$ is the conformal time and $a$ is the scale factor. From the space-space components of Einstein's equations, the vector mode  in the momentum space obeys~\cite{Kahniashvili:2000vm,Paoletti:2008ck,Durrer:1999bk}
\begin{equation}\label{vik}
  V'_i(\tau,\mathbf{k})+2\mathcal{H}V_i(\tau,\mathbf{k})
  =-\frac{16\pi G\,\Pi^{(V)}_i(\mathbf{k})}{a^2\,k}
\end{equation}
in the presence of a vector source $V_i$. Here, $k=|\mathbf{k}|$, $\mathcal{H}$ is the conformal Hubble parameter, $G$ is the Newton's gravitational constant, and $\Pi^{(V)}_i(\mathbf{k})$ is the projected vector component of the ESET.  
The solution to \Eq{vik} is
\begin{equation}\label{vik2}
  V_i(\tau,\mathbf{k})=-\frac{16\pi C G\,\Pi^{(V)}_i(\mathbf{k})}{k},
  \qquad
  C=\frac{\tau}{a^2}.
\end{equation}
For a cosmic background fluid with a constant equation of state $w$, the scale factor evolves as $a\propto \tau^{2/(1+3w)}$, and therefore  $V_i\sim \tau^{\frac{3(-1+w)}{1+3w}}$. Thus, the vector mode decays with the cosmic expansion for $w<1$, but retains  constant during   a post-inflationary stiff or kination phase  with $w=1$.
In this case, the coefficient $C$ is constant and can be written as
    \begin{equation}\label{4}
  C= \mathrm{constant}=\frac{1}{2a_{\mathrm{inf}}^3 H_{\mathrm{inf}}}, \quad \mathrm{if}\ w=1,
\end{equation}
where $H$ is the Hubble parameter and the subscript ``inf'' denotes the time of the end of inflation.  In the second equality in Eq.~(\ref{4}), we have fixed the integration constant by evaluating  $C$  at the end of inflation, which we take  to coincide with the onset of the kination phase.  This non-decaying vector mode can subsequently act as a persistent source of second-order curvature perturbations, which may become large enough to seed PBH formation after horizon reentry.

\textit{Vector-induced Curvature Perturbations.} 
Since the first-order vector mode remains approximately constant during the stiff stage, it can act as an efficient source of higher-order scalar perturbations. We now derive the scalar perturbations  induced  at second order by the first-order vector mode. In the conformal Newtonian gauge, the perturbed spatially flat FLRW metric takes the form
\begin{equation}
  ds^2=a^2\left[-(1+\Phi)d\tau^2+2V_i\,d\tau dx^i+(1-\Psi)\delta_{ij}dx^idx^j\right],
\end{equation}
where $\Phi$ and $\Psi$ denote the second-order scalar perturbations. For simplicity, we assume $\Phi=\Psi$ at second order throughout this work. 

At the second order, the scalar potential $\Phi$ satisfies
\begin{equation}\label{eq6}
  \Phi''+\frac{6(1+w)}{1+3w}\frac{1}{\tau}\Phi'-w\Delta\Phi=S,
\end{equation}
with the source term being 
\begin{equation}
\begin{aligned}
  S
  =&-V_i(2V_i'\mathcal{H}+\Delta V_i)
  -\frac{1}{4}(w-1)\partial_jV_i\partial_iV_j \\
  &-\frac{1}{4}(3+w)\partial_jV_i\partial_jV_i
  -\frac{w\,\Delta V_i\Delta V_i}{12(1+w)\mathcal{H}^2}.
\end{aligned}
\end{equation}
This equation can be derived from the second-order Einstein equations using \texttt{xPand}~\cite{Pitrou:2013hga}.
For the kination stage in which $w=1$, the above equation in momentum space becomes
\begin{equation}
  \Phi_{\mathbf{k}}''+\frac{3}{\tau}\Phi_{\mathbf{k}}'+k^2\Phi_{\mathbf{k}}=S_{\mathbf{k}},
\end{equation}
where
\begin{equation}\label{9}
  \begin{aligned}
    S_{\mathbf{k}}
    = \int \frac{d^3 q}{(2\pi)^{3/2}}
    &
    \Bigg[
      (\mathbf{k}-\mathbf{q})^2
      +\mathbf{q}\!\cdot(\mathbf{k}-\mathbf{q})
      -\frac{\tau^2}{6}\mathbf{q}^2(\mathbf{k}-\mathbf{q})^2
    \Bigg] \\
    &\times
    V_i(\tau,\mathbf{q})V_i(\tau,\mathbf{k}-\mathbf{q}).
  \end{aligned}
\end{equation}
The corresponding retarded  Green-function solution is
\begin{equation}
\label{Phik}
\Phi_k
=
\mathcal A_{\mathbf k}
\left(
1-\frac{2J_1(x)}{x}
\right)
+
\mathcal B_{\mathbf k}
\left(
8-x^2-\frac{16J_1(x)}{x}
\right),
\end{equation}
where  $x\equiv k\tau$, $J_1(x)$ is the first kind Bessel function~\cite{NIST:DLMF}, 
\begin{equation}
\mathcal A_{\mathbf k}
\equiv
\frac{1}{2}
\int \frac{d^3 q}{(2\pi)^{3/2}}
\,V_i(\mathbf q)\,V_i(\mathbf k-\mathbf q),
\end{equation}
and
\begin{equation}
\mathcal B_{\mathbf k}
\equiv\frac{1}{6k^4}
\int \frac{d^3 q}{(2\pi)^{3/2}}
\,\mathbf q^2(\mathbf k-\mathbf q)^2\,
V_i(\mathbf q)\,V_i(\mathbf k-\mathbf q).
\end{equation}
The derivation of this Green-function solution from the general equation valid for arbitrary $w$ is detailed in Appendix~\ref{AA}.

We define the two-point correlator of the vector component of the PMF ESET as~\cite{Mack:2001gc}
\begin{equation}
	\label{eqPi2}
  \left\langle \Pi_{i_1}^{(V)}(\mathbf{k})
  \Pi_{i_2}^{(V)}(\mathbf{k}') \right\rangle=  |\Pi^{(V)}(k)|^2\,P_{i_1 i_2}(\mathbf{k})\,\delta(\mathbf{k}+\mathbf{k}'),
\end{equation}
with the transverse  projection tensor 
$
  P_{ij}=\delta_{ij}-k_i k_j/k^2,
$
and the dimensionless power spectrum of the induced curvature perturbations
\begin{equation}
  \mathcal{P}_{\mathcal R}(k)\,\delta(\mathbf{k}+\mathbf{k}')
  \equiv
  \frac{k^3}{2\pi^2}\frac{16}{9}
  \left\langle
    \Phi_{\mathbf{k}}\Phi_{\mathbf{k}'}
  \right\rangle.
\end{equation}
Introducing the dimensionless variables
\begin{equation}
  u \equiv \frac{|\mathbf{q}|}{k},
  \qquad
  v \equiv \frac{|\mathbf{k}-\mathbf{q}|}{k},
\end{equation}
imposing the initial conditions  well outside the horizon, and evaluating the power spectrum at horizon entry, $k/\mathcal H=1$, i.e. $x=1/2$,  we obtain
\begin{equation}\label{PRK}
  \begin{aligned}
    \mathcal{P}_{\mathcal R}(k)
    &=
\frac{1024\,C^4G^4k^2}{81}
    \int_0^\infty du
    \int_{|1-u|}^{1+u} dv\\
    &\quad\times    
    |\Pi^{(V)}(uk)|^2\,|\Pi^{(V)}(vk)|^2 \\
    &\quad\times
    \frac{1}{u^3v^3}
    \Bigl[
      u^4+(v^2-1)^2+u^2(6v^2-2)
    \Bigr] \\
    &\quad\times
    \Bigl[
      12+31u^2v^2-16(3+8u^2v^2)J_1\!\Bigl(\frac{1}{2}\Bigr)
    \Bigr]^2 .
  \end{aligned}
\end{equation}
The intermediate steps leading to Eq.~(\ref{PRK}) are collected in  Appendix~\ref{AB}. 
During a stiff era, the first-order vector mode remains approximately constant in time, while the Hubble radius continues to grow. 
Consequently, within the kination band, the induced curvature spectrum is enhanced toward smaller $k$, corresponding to larger comoving scales, until the infrared cutoff set by the end of kination is reached. To characterize this scale dependence of the induced spectrum, we define the spectral index
\begin{equation}
  n \equiv \frac{d\ln \mathcal{P}_{\mathcal R}}{d\ln k}.
\end{equation}
An approximate analysis of Eq.~(\ref{PRK}) near the infrared cutoff yields
\begin{equation}\label{Eq18}
  n \simeq -5.
\end{equation}
This result follows from the cutoff-dominated integration region together with the universal kernel structure in Eq.~(\ref{PRK}), and is therefore insensitive to the details of the underlying stress spectrum. The infrared scaling $\mathcal{P}_{\mathcal R}(k)\propto k^{-5}$ can thus be regarded as a characteristic prediction of the present mechanism, and may help distinguish it from other scenarios for enhancing small-scale curvature perturbations. The corresponding derivation of Eq.~(\ref{Eq18}) is presented   in  Appendix~\ref{AD}

\textit{Power spectrum of curvature perturbations.}  
We now consider the case in which  the source of vector perturbations is provided by primordial magnetic fields (PMFs).  The  ESET of  PMFs $\Pi^{(B)}_{mn}(\mathbf{k})$ can be decomposed into scalar, vector, and tensor components~\cite{Durrer:1999bk,Kodama:1985bj,Mukhanov:1990me,Mack:2001gc,Paoletti:2008ck,Yamazaki:2011ys} (see Appendix~\ref{AE}).
The vector component is obtained through the projection~\cite{Mack:2001gc,Kahniashvili:2000vm},
\begin{equation}
\Pi^{(V)}_i(\mathbf{k})
=
P_{in}(\mathbf{k})\,\hat{k}_m\,\Pi^{(B)}_{mn}(\mathbf{k}),
\end{equation}
where 
 $ P_{in}$ is the  transverse projection operator and $\hat{k}_m$ is the unit vector along $\mathbf{k}$.   
 The expression $\Pi^{(B)}_{mn}(\mathbf{k})$ in terms of the magnetic field $\bf{B}$ is given in Eq.~(\ref{E7}) of Appendix~\ref{AE}.   
We assume that the  spectrum of  magnetic field $\bf{B}$, defined in Eq.~(\ref{E8}) of  Appendix~\ref{AE} 
 takes  a power-law form within the finite interval $k_{\rm IR}<k<k_{\rm UV}$
\begin{equation}\label{eq20}
P_B(k)=A_B k^{n_B}\Theta(k_{\rm UV}-k)\Theta(k-k_{\rm IR}),
\end{equation}
where the ultraviolet cutoff $k_{\rm UV}$ and  the infrared cutoff $k_{\rm IR}$ denote, respectively, the largest and smallest comoving wavenumbers over which the magnetic field is generated.
In an inflationary magnetogenesis realization, the ultraviolet cutoff is naturally associated with the shortest mode exiting the horizon at the end of inflation, $k_{\rm UV}\sim a_{\rm inf}H_{\rm inf}$~\cite{Mack:2001gc,Kosowsky:2001xp}.
We also introduce the infrared cutoff $k_{\rm IR}\sim H_{\rm kin}a_{\rm kin}$, where the subscript ``kin'' denotes the end of kination. This cutoff corresponds to the longest mode that reenters the horizon during the stiff epoch~\cite{Bhaumik:2025kuj}. This infrared cutoff is derived  under the sharp-transition approximation, in which the kination phase is assumed to end instantaneously and to be followed by  radiation domination.  This idealized treatment allows us to isolate the effect of the kination epoch while retaining a simple analytical description of the enhanced power spectrum. In a more realistic scenario, however, the transition takes place over a finite duration, during which the  equation-of-state parameter evolves continuously from $w \simeq 1$ to  $w=1/3$. Such a smooth transition modifies both the background dynamics  and the evolution of perturbation modes near the end of kination. Consequently, the characteristic infrared scale $k_\mathrm{IR}$ may be shifted, while the sharp cutoff predicted in the instantaneous-transition approximation may be smoothed out.  Predictions in the vicinity of $k_\mathrm{IR}$ may therefore be sensitive to the detailed dynamics of the kination-to-radiation transition. 

The power-law magnetic spectrum given in Eq.~(\ref{eq20}) is a common  parameterization of stochastic PMFs in CMB analyses~\cite{Mack:2001gc,Paoletti:2010rx,Planck:2015zrl}
and is also naturally motivated by inflationary magnetogenesis
models~\cite{Ratra:1991bn,Martin:2007ue,Subramanian:2015lua,Yokoyama:2015era}.
As a concrete realization, we adopt the Ratra-type magnetogenesis scenario~\cite{Ratra:1991bn,Kobayashi:2019uqs}. In this model, the magnetic spectral index and amplitude are given by
\begin{equation}
\label{eq:nb}
  \begin{aligned}
    n_B&=-2s+3, \\
    A_B&=\frac{16}{\pi}\Gamma\!\left(s-\frac{1}{2}\right)^2
    (a_{\rm inf}H_{\rm inf})^4
    \left(\frac{1}{2a_{\rm inf}H_{\rm inf}}\right)^{-2(s-3)},
  \end{aligned}
\end{equation}
where $s$ is the index of the nonconformal coupling function in the electromagnetic action.
For this class of spectra, the vector stress spectrum appearing in Eq.~(\ref{eqPi2}) can be
approximated as~\cite{Kosowsky:2001xp,Mack:2001gc}
\begin{equation}
    |\Pi^{(V)}(k)|^2
\simeq
    4\pi A_B^2
    \left[
    \frac{n_B\,k^{2n_B+3}}{(n_B+3)(2n_B+3)}
      +\frac{k_{\rm UV}^{2n_B+3}}{2n_B+3}
    \right].
\end{equation}

For a stiff  stage lasting $\Delta N_{\rm kin}$ e-folds, the ultraviolet cutoff can be written as
\begin{equation}
  k_{\rm UV}
  =H_{\rm inf}
  \left(\frac{H_0}{H_{\rm eq}}\right)^{2/3}
  \left(\frac{H_{\rm eq}}{H_{\rm inf}}\right)^{1/2}
  \exp\!\left(\frac{3w-1}{4}\Delta N_{\rm kin}\right),
\end{equation}
where $H_0\sim10^{-42}\,{\rm GeV}$ is the present Hubble scale and $H_{\rm eq}\sim1.5\times10^{-37}\,{\rm GeV}$ the Hubble scale at matter-radiation equality~\cite{Tomberg:2021ajh}. The corresponding infrared cutoff is given by~\cite{Bhaumik:2025kuj}
\begin{equation}
  k_{\rm IR}
  =(H_{\rm inf}H_{\rm eq})^{1/2}
  \left(\frac{H_0}{H_{\rm eq}}\right)^{2/3}
  \exp\!\left(-\frac{3w+3}{4}\Delta N_{\rm kin}\right).
\end{equation}

We choose the representative parameter values
\((H_{\rm inf},\Delta N_{\rm kin},s)=(10^{14.31}\,{\rm GeV},9,2)\), which specify the post-inflationary evolution and the power-law magnetic spectrum. 
For this  choice, one obtaines 
\begin{equation}\label{24}
k_{\rm UV}\simeq 2.8\times10^{25}\,{\rm Mpc}^{-1},
\qquad
k_{\rm IR}\simeq 4.2\times10^{17}\,{\rm Mpc}^{-1}. 
\end{equation}
The Ratra-type spectrum then  yields
\begin{eqnarray}
n_B=-1,
\qquad
A_B\simeq 7.6\times10^{50}\,{\rm Mpc}^{-2}.
\end{eqnarray}
This  corresponds to a finite-band magnetic amplitude of order \(B_{\rm rms}\sim O(10^{2})\,{\rm nG}\). This magnetic amplitude is consistent with the BBN bounds~\cite{Kawasaki:2012va}  and with the constraints from the PTA observations~\cite{Saga:2018qrx} on the total magnetic energy density. Although $B_{\rm rms}$ exceeds the limits inferred  from CMB  observations~\cite{Jedamzik:1999bm,Planck:2015zrl},  this does not directly exclude the parameter choice considered here, because the CMB constraints apply mainly to much larger comoving scales, approximately 400\,{\rm pc}$\sim\mathcal{O}(1)\,${\rm Mpc}, whereas the magnetic field in our scenario is generated on the much smaller scales specified in Eq.~(\ref{24}).

\begin{figure}[t]
  \centering
  \includegraphics[width=0.48\textwidth]{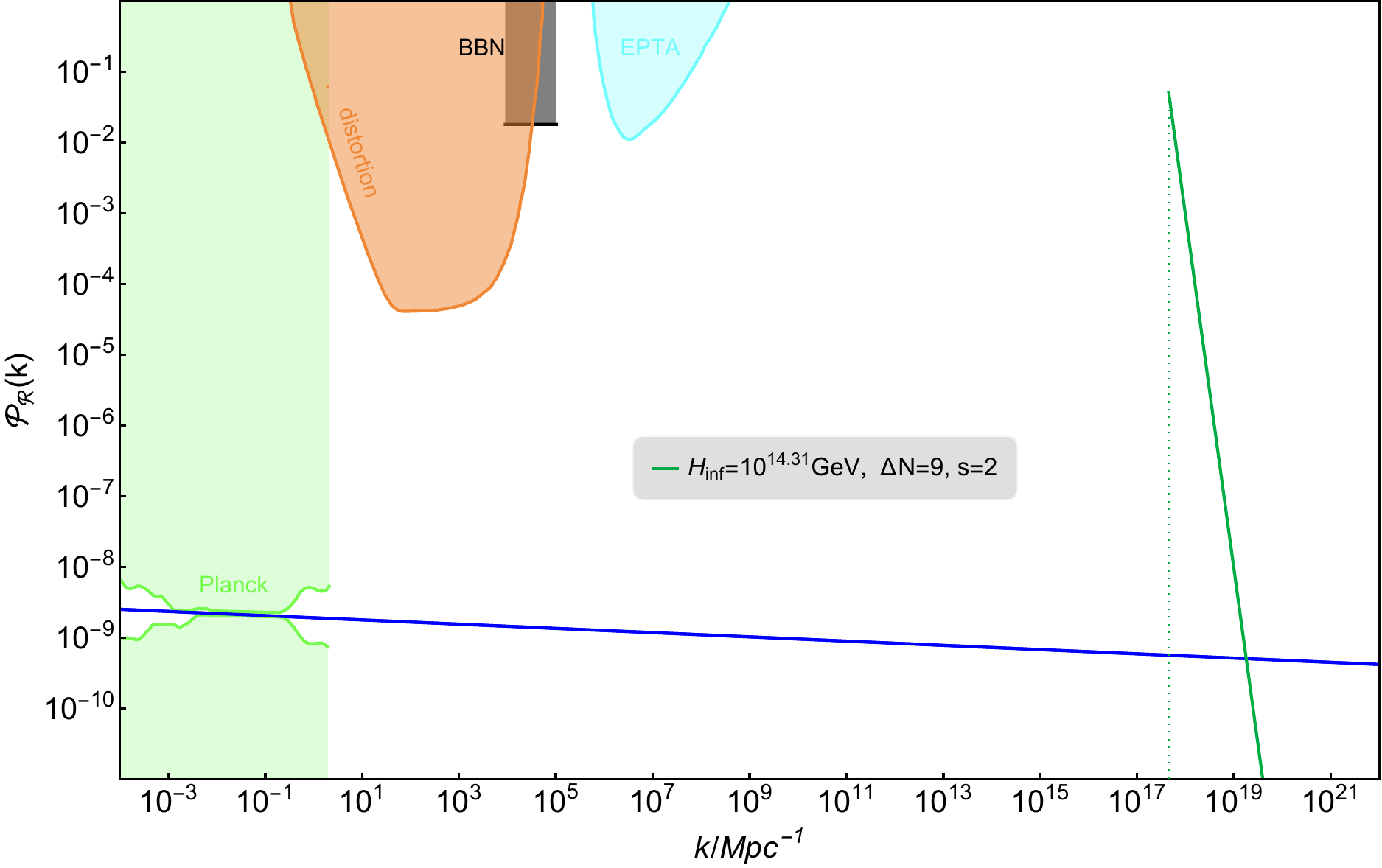}
\caption{The green solid curve shows the vector-induced curvature power spectrum $\mathcal{P}_{\mathcal R}(k)$ obtained from the PMFs with parameters being $(H_{\rm inf},\Delta N_{\rm kin},s)=(10^{14.31}\,{\rm GeV},9,2)$, while the dotted vertical line marks the corresponding infrared cutoff scale $k_{\rm IR}$. The blue solid line denotes the  nearly scale-invariant power spectrum allowed by CMB.  The spectrum can reach \(O(10^{-2})\)   and exhibits the approximate scaling $\mathcal{P}_{\mathcal R}(k)\propto k^{-5}$. The shaded regions show current observational upper bounds from Planck~\cite{Planck:2018vyg}, BBN~\cite{Inomata:2016uip}, EPTA~\cite{Inomata:2018epa}, and $\mu$-distortion of CMB~\cite{Fixsen:1996nj}.}
  \label{fig1}
\end{figure}

\Fig{fig1} shows the power spectrum of the induced curvature perturbations  obtained from Eq.~(\ref{PRK}) for the parameter choice $(H_{\rm inf},\Delta N_{\rm kin},s)=(10^{14.31}\,{\rm GeV},9,2)$. The green solid curve exhibits the characteristic behavior $\mathcal{P}_{\mathcal R}(k)\propto k^{-5}$,  consistent with the universal infrared scaling derived from  the vector-induced kernel. The sharp feature at $k=k_{\rm IR}$ is a direct consequence of imposing the infrared cutoff in the momentum integration. The $k^{-5}$ behavior, by contrast, is obtained within the idealized pure-kination dynamics with $w=1$. In a realistic scenario, the kination-to-radiation transition occurs over a finite duration, during which $w$ evolves continuously from $1$ to $1/3$. The resulting departure from pure kination causes the vector perturbations to decay, thereby suppressing the source of the induced curvature perturbations. Therefore, the spectrum is expected to deviate from the  $k^{-5}$ behavior for modes sensitive to the transition, with the largest modifications occurring near the infrared end of the spectrum. Modes evolving sufficiently deep in the kination era, on the other hand, remain largely insensitive to the transition details and approximately preserve the  pure-kination $k^{-5}$ scaling.

The spectral amplitude is significantly enhanced around the infrared cutoff scale $k_{\rm IR}$ relative to the CMB amplitude. This enhancement in curvature perturbations can efficiently seed the formation of PBHs.

\textit{PBHs.} 
The present PBH abundance relative to dark matter is defined as~\cite{Young:2014ana,Carr:1975qj}
\begin{equation}
\frac{\Omega_{\rm PBH}}{\Omega_{\rm DM}}
  =\int dM\,f(M),
\end{equation} 
where \(f(M)\) denotes the present-day PBH mass function.   In the present scenario, the vector-induced scalar perturbations may exhibit nonstandard subhorizon evolution. In addition, magnetic pressure can increase the effective collapse threshold in magnetized overdense regions, thereby suppressing PBH formation~\cite{Kushwaha:2024zhd}. 

More generally, the conditions for PBH formation should be modified by various effects, such as non-ideal fluid corrections and high-energy particle collisions in the early Universe~\cite{Saini:2017tsz,Yuwen:2026hxu}.  A smooth transition may also modify the curvature spectrum near $k_{\rm IR}$, the relation between comoving scale and PBH mass, and the collapse threshold, thereby changing the PBH mass spectrum in the corresponding mass range. A fully consistent treatment of these effects requires a dedicated analysis. In this work, we therefore adopt the conventional Press-Schechter approach~\cite{Bhattacharya:2023ztw,Press:1973iz,Carr:1975qj, Green:1997sz,Green:2004wb,Sasaki:2018dmp}  as a first estimate of the PBH abundance, leaving a more complete investigation to future work.

\begin{figure}[t]
  \centering
  \includegraphics[width=0.48\textwidth]{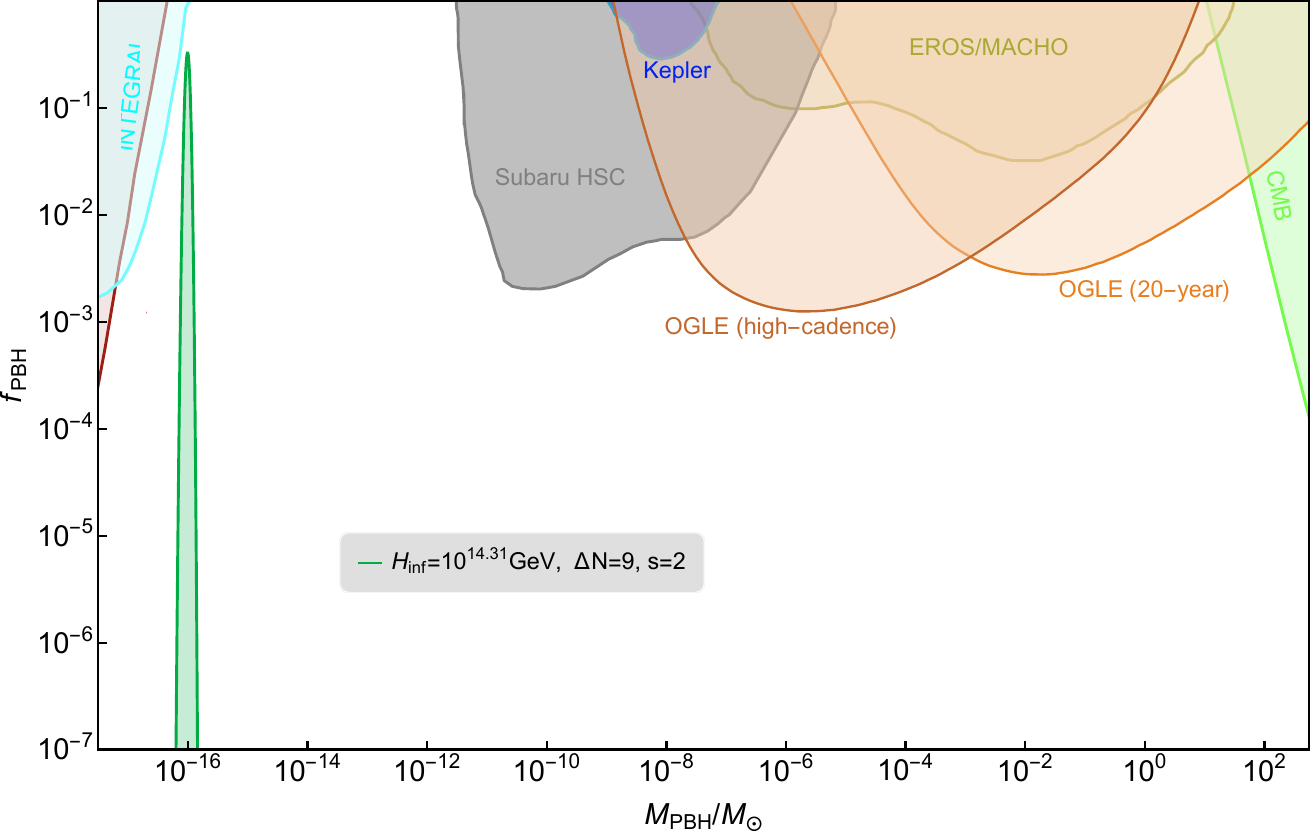}
  \caption{Present-day PBH mass function $f(M)$ with  the parameter choice $(H_{\rm inf},\Delta N_{\rm kin},s)=(10^{14.31}\,{\rm GeV},9,2)$. 
  The shaded regions show current observational bounds on the PBH abundance from the extragalactic $\gamma$-ray background (EG $\gamma$)~\cite{Carr:2020gox}, INTEGRAL~\cite{Laha:2019ssq}, microlensing surveys (Kepler, Subaru/HSC, EROS/MACHO, and OGLE)~\cite{Griest:2013aaa,Griest:2013esa,Niikura:2017zjd,EROS-2:2006ryy, Mroz2024OGLE20Year,Mroz2024OGLEHighCadence}, and the cosmic microwave background (CMB)~\cite{Poulin:2017bwe} }
  \label{fig3}
\end{figure}

\Fig{fig3} displays the resulting $f(M)$ for the representative parameter choice, $(H_{\rm inf},\Delta N_{\rm kin},s)=(10^{14.31}\,{\rm GeV},9,2)$, along with some observational constraints on PBH dark matter. The resulting PBH abundance is compatible with the displayed observational bounds and can account for a significant fraction of the dark matter.

\textit{Discussions and conclusions.}
We have demonstrated that  vector modes soured  by the vector component of  the electromagnetic stress-energy tensor of primordial magnetic fields   can induce significant curvature perturbations at second order during a stiff post-inflationary epoch.  Since the first-order vector mode remains approximately constant during kination, it acts as a persistent nonlinear source for scalar perturbations. The resulting curvature power spectrum is enhanced toward the infrared cutoff and follows the characteristic scaling \(\mathcal P_{\mathcal R}\propto k^{-5}\). When these enhanced curvature perturbations reenter the horizon during the kination phase, they can produce sufficiently large  overdensities to gravitationally collapse   into PBHs.  Applying this mechanism to PMFs produced in a Ratra-type magnetogenesis scenario, we find that the resulting PBHs can account for a substantial fraction of the dark matter.

This work proposes a new mechanism for enhancing curvature perturbations and producing PBHs. In contrast to many conventional PBH scenarios, where the relevant enhancement occurs directly in the first-order scalar sector, the present mechanism relies on second-order scalar perturbations induced by magnetically sourced vector modes through nonlinear gravitational coupling. The characteristic infrared scaling of the induced curvature spectrum provides a distinctive feature of this scenario.

Several issues deserve further investigation. First, the PBH abundance estimate presented here is based on the conventional Press-Schechter formalism. A more complete treatment should include the nonstandard subhorizon evolution of the vector-induced scalar perturbations, as well as the possible modification of the collapse threshold by magnetic pressure.  Second, our  results are  obtained under the idealized assumption of a pure-kination phase followed by a sharp transition to radiation domination. They are expected to be most reliable for modes whose relevant evolution takes place sufficiently deep inside the $w\simeq1$ epoch. A quantitative assessment of how a realistic kination-to-radiation transition modifies the curvature power spectrum and the resulting PBH mass function requires specifying the transition dynamics and numerically evolving the perturbations through the time-dependent background. Third,  because the curvature perturbations are generated at second order from a quadratic vector source, their statistics are expected to be intrinsically non-Gaussian. A dedicated analysis of this non-Gaussianity is therefore essential for a more accurate prediction of the PBH abundance.
  Finally, the same enhanced curvature perturbations may source a stochastic gravitational-wave background, whose spectrum could differ from standard scalar-induced gravitational waves because the curvature perturbations obey a sourced second-order evolution equation during their amplification. We leave these questions for future work.


\begin{acknowledgments}
  This work was supported by  the National Natural Science Foundation of China (Grants No.~12275080, No.~12405056,  and No.~12635002), the Natural Science Foundation of Hunan Province (Grant No.~2025JJ40006), and the Innovative Research Group of Hunan Province (Grant No.~2024JJ1006).
\end{acknowledgments}

%
\appendix
\onecolumngrid
\begingroup
\small
\setlength{\parindent}{2em}
\setlength{\parskip}{0.2em}
\renewcommand{\baselinestretch}{1.05}\selectfont
\newpage

\section{The Green-function solution for metric scalar perturbations}
\label{AA}
From Eq.~(6), we obtain that in the momentum space,   the metric scalar perturbation $\Phi_k$ satisfies the following  equation
\begin{equation}\label{A1}
	\Phi_k''+\frac{6(1+w)}{1+3w}\frac{1}{\tau}\Phi_k'
+wk^2\Phi_k=S_k(\tau).
\end{equation}
Here $S_k$ is the source term.  To solve the above equation, we first consider the corresponding homogeneous equation,
\begin{equation}
	\Phi_k''+\frac{6(1+w)}{1+3w}\frac{1}{\tau}\Phi_k'
	+wk^2\Phi_k=0, 
\end{equation}
which has  two independent specific solutions $u_1$ and $u_2$: 
\begin{equation}\label{A3}
	u_1(\tau)=\tau^{-\nu}J_\nu(\sqrt{w}\,k\tau),
	\qquad
	u_2(\tau)=\tau^{-\nu}Y_\nu(\sqrt{w}\,k\tau),
\end{equation}
where $J_\nu$ and $Y_\nu$ are the Bessel functions of the first and second
kinds, respectively, and
\begin{equation}
	\nu=\frac{1}{2}\frac{5+3w}{1+3w}.
\end{equation}
The corresponding Wronskian is
\begin{equation}
	W(\tau)=u_1u_2'-u_1'u_2=\frac{2}{\pi\tau^{2\nu+1}}.
\end{equation}

For a second-order linear differential equation, the retarded Green's function can be expressed  in terms of the two independent specific  solutions  and their Wronskian, taking the form:  
\begin{equation}
	\begin{aligned}
		G(\tau,\bar{\tau})
		&=\frac{u_1(\bar{\tau})u_2(\tau)-u_1(\tau)u_2(\bar{\tau})}
		{W(\bar{\tau})}
		\\
		&=\frac{\pi}{2}\frac{\bar{\tau}^{\,\nu+1}}{\tau^\nu}
		\left[
		J_\nu(\sqrt{w}\,k\bar{\tau})Y_\nu(\sqrt{w}\,k\tau)
		-
		J_\nu(\sqrt{w}\,k\tau)Y_\nu(\sqrt{w}\,k\bar{\tau})
		\right].
	\end{aligned}
\end{equation}
Using the Green-function method, we obtain the  solution of Eq.~(\ref{A1}) 
\begin{equation}\label{A7}
	\Phi_k(\tau)= \int_{\tau_i}^{\tau}d\bar{\tau}\,
	G(\tau,\bar{\tau})\,S_k(\bar{\tau}).
\end{equation}

During  the kination era, which has $w=1$, we have $\nu=1$.  Then Eq.~(\ref{A7})
can be  reduced to
\begin{equation}
	\Phi_k(\tau)=\frac{\pi}{2\tau}\int_{\tau_i}^{\tau}d\bar{\tau}\,
	\bar{\tau}^{2}
	\left[
	J_1(k\bar{\tau})Y_1(k\tau)
	-
	J_1(k\tau)Y_1(k\bar{\tau})
	\right]S_k(\bar{\tau}),
\end{equation}
where
\begin{equation}\label{A9}
	\begin{aligned}
	S_{\mathbf{k}}(\bar{\tau})
		= \int \frac{d^3 q}{(2\pi)^{3/2}}
		&\Big[
		(\mathbf{k}-\mathbf{q})^2
		+\mathbf{q}\!\cdot(\mathbf{k}-\mathbf{q})
		-\frac{\bar{\tau}^2}{6}\mathbf{q}^2(\mathbf{k}-\mathbf{q})^2
		\Big] 
		V_i(\mathbf{q})V_i(\mathbf{k}-\mathbf{q}),
	\end{aligned}
\end{equation}
as given in Eq.~(9).

We choose the initial time such that the relevant modes are well outside the
horizon, $x_i=k\tau_i\ll1$. In this limit, the lower bound can be taken as
$x_i\to0$, and the remaining time integrals can be evaluated analytically
using standard identities of Bessel functions~\cite{NIST:DLMF}. Using the following relations for Bessel functions
\begin{equation}
	\begin{aligned}
		\int_0^x d\bar{x}\,\bar{x}^2
		\left[J_1(\bar{x})Y_1(x)-J_1(x)Y_1(\bar{x})\right]
		&=\frac{2}{\pi}\Big[x-2J_1(x)\Big],\\
		\int_0^x d\bar{x}\,\bar{x}^4
		\left[J_1(\bar{x})Y_1(x)-J_1(x)Y_1(\bar{x})\right]
		&=\frac{2}{\pi}\Big[x^3-8x+16J_1(x)\Big],
	\end{aligned}
\end{equation}
we obtain that the Green-function solution can be re-expressed as 
\begin{equation}\label{A11}
	\begin{aligned}
	\Phi_{\mathbf k}(x)
		=
		\int \frac{d^3q}{(2\pi)^{3/2}}
		\left[
	\frac{(\mathbf k-\mathbf q)^2+\mathbf q\cdot(\mathbf k-\mathbf q)}
		{k^2}
		\left(
		1-\frac{2J_1(x)}{x}
		\right)
		-\frac{\mathbf q^2(\mathbf k-\mathbf q)^2}{6k^4}
		\left(
		x^2-8+\frac{16J_1(x)}{x}
		\right)
		\right]V_i(\mathbf q)V_i(\mathbf k-\mathbf q).
	\end{aligned}
\end{equation}
Here $x\equiv k\tau$. 
By taking the symmetry  under
$\mathbf q\leftrightarrow \mathbf k-\mathbf q$,   the Green-function solution in  Eq.~(\ref{A11}) can be further simplified to:   
  \begin{equation}
	\Phi_k (x)
	=
	\mathcal A_{\mathbf k}
	\left(
	1-\frac{2J_1(x)}{x}
	\right)
	+
	\mathcal B_{\mathbf k}
	\left(
	8-x^2-\frac{16J_1(x)}{x}
	\right),
\end{equation}
where 
\begin{equation}
	\mathcal A_{\mathbf k}
	\equiv
	\frac{1}{2}
	\int \frac{d^3 q}{(2\pi)^{3/2}}
	\,V_i(\mathbf q)\,V_i(\mathbf k-\mathbf q),
\end{equation}
and
\begin{equation}
	\mathcal B_{\mathbf k}
	\equiv
	\frac{1}{6k^4}
	\int \frac{d^3 q}{(2\pi)^{3/2}}
	\,\mathbf q^2(\mathbf k-\mathbf q)^2\,
	V_i(\mathbf q)\,V_i(\mathbf k-\mathbf q).
\end{equation}
 

\section{Power spectrum of  scalar perturbations}
\label{AB}

 The power spectrum of metric scalar perturbation $\Phi_\mathbf {k}$ is defined as 
\begin{equation}\label{BC1}
	\mathcal P_\Phi(k)\,\delta(\mathbf{k}+\mathbf{k}')
	=\frac{k^3}{2\pi^2}\left\langle \Phi_{\mathbf k}  \Phi_{\mathbf k'}  \right\rangle .
\end{equation}
Here,  $\left\langle \Phi_{\mathbf k}  \Phi_{\mathbf k'}  \right\rangle$ is the  two-point correlation function of  $\Phi_{\mathbf k}$. 
Using the Green-function solution of $\Phi_k$ obtained in the above section~\ref{AA},  we find that this two-point correlation function  can be expressed as 
\begin{equation} \label{B1}
	\begin{aligned}
		\left\langle \Phi_{\mathbf k} \Phi_{\mathbf k'}  \right\rangle
		=\frac{\pi^2}{4\tau^2}
		\int_{\tau_i}^{\tau} d\bar{\tau}_1
		\int_{\tau_i}^{\tau} d\bar{\tau}_2\,
		\bar{\tau}_1^2 \bar{\tau}_2^2\,
		\mathcal{G}(k;\bar{\tau}_1,\tau)\,
		\mathcal{G}(k';\bar{\tau}_2,\tau)\,
		\left\langle
		S_{\mathbf k} (\bar{\tau}_1)S_{\mathbf k'}(\bar{\tau}_2)		\right\rangle ,
	\end{aligned}
\end{equation}
where
\begin{equation}
	\mathcal{G}(k;\bar{\tau},\tau)
	\equiv
	J_1(k\bar{\tau})Y_1(k\tau)-J_1(k\tau)Y_1(k\bar{\tau}) .
\end{equation}

Substituting the solution of the first-order vector perturbation,	$V_i(\mathbf{k})
	=
	-\frac{16\pi C G}{k}\,\Pi_i^{(V)}(\mathbf{k})$, 
into Eq.~(\ref{A9}) yields the two-point correlation of source term appearing in Eq.~(\ref{B1})  
\begin{equation}\label{eq:Sksk_preWick}
	\begin{aligned}
		\left\langle
		S_{\mathbf k} (\bar{\tau}_1)\,S_{\mathbf k'} (\bar{\tau}_2)
		\right\rangle
		=
		\int \frac{d^3 q_1}{(2\pi)^{3/2}}
		\int \frac{d^3 q_2}{(2\pi)^{3/2}}
		&\Bigg[
		(\mathbf{k}-\mathbf{q}_1)^2
		+\mathbf{q}_1\!\cdot(\mathbf{k}-\mathbf{q}_1)
		-\frac{\bar{\tau}_1^2}{6}\,
		\mathbf{q}_1^2(\mathbf{k}-\mathbf{q}_1)^2
		\Bigg]
		\\
		\times&
		\Bigg[
		(\mathbf{k}'-\mathbf{q}_2)^2
		+\mathbf{q}_2\!\cdot(\mathbf{k}'-\mathbf{q}_2)
		-\frac{\bar{\tau}_2^2}{6}\,
		\mathbf{q}_2^2(\mathbf{k}'-\mathbf{q}_2)^2
		\Bigg]
		\\
		\times&
		\frac{(16\pi C G)^4}
		{|\mathbf{q}_1|\,|\mathbf{q}_2|\,
			|\mathbf{k}-\mathbf{q}_1|\,|\mathbf{k}'-\mathbf{q}_2|}
		\\
		\times&
		\left\langle
		\Pi_{i_1}^{(V)}(\mathbf{q}_1)\,
		\Pi_{i_1}^{(V)}(\mathbf{k}-\mathbf{q}_1)\,
		\Pi_{i_2}^{(V)}(\mathbf{q}_2)\,
		\Pi_{i_2}^{(V)}(\mathbf{k}'-\mathbf{q}_2)
		\right\rangle .
	\end{aligned}
\end{equation}
 According to the Wick's     theorem,
Eq.~(\ref{eq:Sksk_preWick}) can be re-written as 
\begin{equation}\label{B6}
	\begin{aligned}
		\left\langle
		S_{\mathbf k} (\bar{\tau}_1)S_{\mathbf k'} (\bar{\tau}_2)
		\right\rangle
		=\delta(\mathbf{k}+\mathbf{k}')
		\int \frac{d^3 q}{(2\pi)^3}\,
		&\Bigg[
		(\mathbf{k}-\mathbf{q})^2
		+\mathbf{q}\cdot(\mathbf{k}-\mathbf{q})
		-\frac{\bar{\tau}_1^2}{6}\,
		q^2|\mathbf{k}-\mathbf{q}|^2
		\Bigg]
		\\
		\times\,
		&\Bigg[
		1+\frac{\bigl[\mathbf{q}\cdot(\mathbf{k}-\mathbf{q})\bigr]^2}
		{q^2|\mathbf{k}-\mathbf{q}|^2}
		\Bigg]
		\frac{(16\pi C G)^4}{q\,|\mathbf{k}-\mathbf{q}|}
		|\Pi^{(V)}(q)|^2\,
		|\Pi^{(V)}(|\mathbf{k}-\mathbf{q}|)|^2
		\\
		\times\,
		&\Bigg[
		\frac{k^2}{q\,|\mathbf{k}-\mathbf{q}|}
		-\frac{\bar{\tau}_2^2}{3}\,
		q\,|\mathbf{k}-\mathbf{q}|
		\Bigg].
	\end{aligned}
\end{equation}



Substituting  Eq.~(\ref{B6}) into Eq.~(\ref{B1} ) and using the definition of power spectrum given in Eq.~(\ref{BC1}), we get the expression of  power spectrum $\mathcal P_\Phi(k)$:
\begin{equation}\label{C2}
	\begin{aligned}
		\mathcal P_\Phi(k)
		= \frac{k^3}{8\tau^2}
		\int_{\tau_i}^{\tau} d\bar{\tau}_1
		\int_{\tau_i}^{\tau} d\bar{\tau}_2\,
		\bar{\tau}_1^2\bar{\tau}_2^2\,
		\mathcal{G}(k;\bar{\tau}_1,\tau)\,
		\mathcal{G}(k;\bar{\tau}_2,\tau)
		\int \frac{d^3 q}{(2\pi)^3}\,
		\mathcal{I}(\mathbf{k},\mathbf{q},\mathbf{p};\bar{\tau}_1,\bar{\tau}_2), 
	\end{aligned}
\end{equation}
where 
$\mathbf{p}=\mathbf{k}-\mathbf{q}$, and 
\begin{equation}
	\begin{aligned}
		\mathcal{I}(\mathbf{k},\mathbf{q},\mathbf{p};\bar{\tau}_1,\bar{\tau}_2)
		\equiv\,
		&\left[
		p^2+\mathbf{q}\cdot\mathbf{p}
		-\frac{\bar{\tau}_1^2}{6}\,q^2 p^2
		\right]
		\left[
		1+\frac{(\mathbf{q}\cdot\mathbf{p})^2}{q^2 p^2}
		\right]
		\left[
		\frac{k^2}{q^2 p^2}-\frac{\bar{\tau}_2^2}{3}
		\right]
		(16\pi C G)^4
		|\Pi^{(V)}(q)|^2 |\Pi^{(V)}(p)|^2 
	\end{aligned}
\end{equation}
is the kernel function. 
From $\mathbf{q}\cdot\mathbf{p}=\frac{1}{2}(\mathbf{q}\cdot\mathbf{p}+\mathbf{p}\cdot\mathbf{q})$, one can derive 
$
\mathbf{q}\cdot\mathbf{p}=\mathbf{q}\cdot\mathbf{k}-q^2
=\frac{k^2-q^2-p^2}{2},
$
which leads to $\mathbf{q}\cdot\mathbf{k}=\frac{k^2+q^2-p^2}{2}$.
 Without loss of generality, we can choose
$\mathbf k$ along the $z$ axis and then obtain $$\mathbf{q}\cdot\mathbf{k}=qk\cos\theta.$$
Using the above relations, we have \begin{eqnarray}\label{C4}
d^3q=2\pi q^2dq\,d\cos\theta
=\frac{2\pi q p}{k}\,dq\,dp 
\end{eqnarray}
with $p\in[|k-q|,k+q]$. Substituting Eq.~(\ref{C4}) into Eq.~(\ref{C2}) gives
\begin{equation}\label{C5}
	\begin{aligned}
		\mathcal P_\Phi(k)
		= \frac{k^3}{8\tau^2}
		\int_{\tau_i}^{\tau} d\bar{\tau}_1
		\int_{\tau_i}^{\tau} d\bar{\tau}_2\,
		\bar{\tau}_1^2\bar{\tau}_2^2\,
		\mathcal{G}(k;\bar{\tau}_1,\tau)\,
		\mathcal{G}(k;\bar{\tau}_2,\tau)\,
		\frac{1}{(2\pi)^2}
		\int_0^\infty dq
		\int_{|k-q|}^{k+q}\frac{qp}{k}\,dp\,
		\mathcal{F}(k,q,p;\bar{\tau}_1,\bar{\tau}_2),
	\end{aligned}
\end{equation}
with
\begin{eqnarray}
	\mathcal{F}(k,q,p;\bar{\tau}_1,\bar{\tau}_2)		&\equiv&\,
		\left[
		1+\frac{(k^2-q^2-p^2)^2}{4q^2p^2}
		\right]
		\\
		&&\times
		\left[
		\frac{k^2(k^2+p^2-q^2)}{2p^2q^2}
		-\frac{k^2}{6}\bar{\tau}_1^2
		-\frac{k^2+p^2-q^2}{6}\bar{\tau}_2^2
		+\frac{p^2q^2}{18}\bar{\tau}_1^2\bar{\tau}_2^2
		\right]
		\\
		&&\times (16\pi C G)^4
		|\Pi^{(V)}(q)|^2 |\Pi^{(V)}(p)|^2 .
\end{eqnarray}

Introducing the  following dimensionless
variables, 
\begin{equation}
	u=\frac{q}{k},
	\qquad
	v=\frac{p}{k},
	\qquad
	\bar{x}_1=k\bar{\tau}_1,
	\qquad
	\bar{x}_2=k\bar{\tau}_2,
	\qquad
	x=k\tau,
\end{equation}
we obtain that Eq.~(\ref{C5}) can be re-expressed as 
\begin{equation}
	\begin{aligned}
		\mathcal P_\Phi(k)
		= \frac{k^2}{8x^2}
		\int_{x_i}^{x} d\bar{x}_1
		\int_{x_i}^{x} d\bar{x}_2\,
		\bar{x}_1^2\bar{x}_2^2\,
		\mathcal{G}(\bar{x}_1,x)\,
		\mathcal{G}(\bar{x}_2,x)\,
		\frac{1}{(2\pi)^2}
		\int_0^\infty du
		\int_{|1-u|}^{1+u} dv\,
		\bar{\mathcal{M}}(u,v;\bar{x}_1,\bar{x}_2),
	\end{aligned}
\end{equation}
where
\begin{eqnarray}
	\mathcal{G}(\bar{x},x)
		\equiv J_1(\bar{x})Y_1(x)-J_1(x)Y_1(\bar{x}), 
\end{eqnarray}
and
\begin{eqnarray}\bar{\mathcal{M}} (u,v;\bar{x}_1,\bar{x}_2)
		&\equiv& uv
		\left[
		1+\frac{(-1+u^2+v^2)^2}{4u^2v^2}
		\right]
		\left[
		\frac{1+v^2-u^2}{2u^2v^2}
		-\frac{\bar{x}_1^2}{6}
		-\frac{1+v^2-u^2}{6}\bar{x}_2^2
		+\frac{u^2v^2}{18}\bar{x}_1^2\bar{x}_2^2
		\right]
\nonumber		\\		&&\times (16\pi C G)^4
		|\Pi^{(V)}(uk)|^2 |\Pi^{(V)}(vk)|^2 
\end{eqnarray}
is the dimensionless kernel function. 

Since the internal momenta $q$ and $p=|\mathbf k-\mathbf q|$ appear symmetrically in  the original
convolution,   the final momentum integral should remain
invariant under the exchange $u\leftrightarrow v$. To make this symmetry explicit, we define a symmetrized kernel as follows:
\begin{equation}
	\mathcal{M}(u,v;\bar{x}_1,\bar{x}_2)
	\equiv \frac{1}{2}\Big[ \bar{\mathcal{M}}(u,v;\bar{x}_1,\bar{x}_2)
	+\bar{\mathcal{M}}(v, u;\bar{x}_1,\bar{x}_2)\Big],
\end{equation}
and then achieve  
\begin{eqnarray}
	\mathcal{M}(u,v;\bar{x}_1,\bar{x}_2)
		&=& uv
		\left[
		1+\frac{(-1+u^2+v^2)^2}{4u^2v^2}
		\right]
		\left[
		\frac{1}{2u^2v^2}
		-\frac{\bar{x}_1^2+\bar{x}_2^2}{6}
		+\frac{u^2v^2}{18}\bar{x}_1^2\bar{x}_2^2
		\right]
		\\
		&&\times (16\pi C G)^4
		|\Pi^{(V)}(uk)|^2 |\Pi^{(V)}(vk)|^2 .
\end{eqnarray}
Thus, the power spectrum can be further re-written as
\begin{equation}
	\begin{aligned}
		\mathcal P_\Phi(k)
		= \frac{k^2}{8x^2}
		\int_{x_i}^{x} d\bar{x}_1
		\int_{x_i}^{x} d\bar{x}_2\,
		\bar{x}_1^2\bar{x}_2^2\,
		\mathcal{G}(\bar{x}_1,x)\,
		\mathcal{G}(\bar{x}_2,x)\,
		\frac{1}{(2\pi)^2}
		\int_0^\infty du
		\int_{|1-u|}^{1+u} dv\,
		\mathcal{M}(u,v;\bar{x}_1,\bar{x}_2).
	\end{aligned}\label{C13}
\end{equation}

Now, we consider the initial scalar mode to be well outside the horizon, which indicates $x_i\to0$. We then take the $\bar{x}_1$ and $\bar{x}_2$ integrals in Eq.~(\ref{C13}) to obtain 
\begin{equation}\label{B18}
	\begin{aligned}
		\mathcal P_\Phi(k)
		= 2
		(8CG)^4 k^2
		\int_0^\infty du
		\int_{|1-u|}^{1+u} dv\,
		|\Pi^{(V)}(uk)|^2 |\Pi^{(V)}(vk)|^2\,
		\mathcal{K}(u,v;x),
	\end{aligned}
\end{equation}
where
\begin{equation}
	\begin{aligned}
		\mathcal{K}(u,v;x)
		\equiv\,
	&\frac{uv}{x^2}
		\left[
		1+\frac{(-1+u^2+v^2)^2}{4u^2v^2}
		\right]
		\\&\ \times
		\left[
		\frac{1}{2u^2v^2}\Big(x-2J_1(x)\Big)^2
		-\frac{1}{3}\Big(x-2J_1(x)\Big)\Big(x^3-8x+16J_1(x)\Big)
		+\frac{u^2v^2}{18}\Big(x^3-8x+16J_1(x)\Big)^2
		\right].
	\end{aligned}
\end{equation}

The induced scalar perturbation relevant for PBH formation is evaluated when the corresponding mode reenters the horizon. Since $\mathcal H=1/(2\tau)$ during the stiff epoch,
  the horizon-entry condition becomes 
$k/\mathcal H=1$, which means that  \begin{equation}
x=k\tau=\frac{k}{2\mathcal H}=\frac{1}{2}.
\end{equation}
Considering this condition in Eq.~(\ref{B18}), we have 
\begin{equation}
	\begin{aligned}
		\mathcal P_\Phi(k)
		=\frac{64 C^4 G^4 k^2}{9}
		\int_0^\infty du
		\int_{|1-u|}^{1+u} dv\,
		|\Pi^{(V)}(uk)|^2 |\Pi^{(V)}(vk)|^2\,
		\mathcal{T}(u,v),
	\end{aligned}
\end{equation}
with
\begin{equation}
	\mathcal{T}(u,v)
		\equiv
		\frac{1}{u^3v^3}
		\left[
		u^4+(v^2-1)^2+u^2(6v^2-2)
		\right]
		\left[	12+31u^2v^2-16(3+8u^2v^2)J_1\!\left(\frac12\right)
		\right]^2 .
\end{equation}

Finally, we convert the power spectrum of  metric scalar perturbations $\Phi$ to that of   curvature
perturbations $\mathcal R$. In the Newtonian gauge, we assume that   $\Phi$ and $\mathcal R$  satisfy the relation
\begin{equation}
	\mathcal R\simeq\frac{5+3w}{3(1+w)}\Phi.
\end{equation}
When $w=1$, one has 
\begin{equation}
	\mathcal R=\frac{4}{3}\Phi,
\end{equation}
which indicates that $
	\mathcal P_{\mathcal R}(k)=\frac{16}{9}\mathcal P_\Phi(k)$. Thus, the power spectrum of curvature perturbations takes the form 
\begin{equation}
	\begin{aligned}
		\mathcal P_{\mathcal R}(k)
		=\frac{1024 C^4 G^4 k^2}{81}
		\int_0^\infty du
		\int_{|1-u|}^{1+u} dv\,
		|\Pi^{(V)}(uk)|^2 |\Pi^{(V)}(vk)|^2\,
		\mathcal T(u,v).
	\end{aligned}
\end{equation}


\section{Infrared spectral index of the  power spectrum}
\label{AD}

We consider a finite comoving band   bounded by the ultraviolet cutoff ($k_{\rm UV}$) and the infrared cutoff ($k_{\rm IR}$). These cutoffs arise from the finite range of modes that reenter the horizon during the stiff epoch. The ultraviolet cutoff is associated with the shortest mode exiting the horizon at the end of inflation, $k_{\rm UV}\sim a_{\rm inf}H_{\rm inf}$~\cite{Mack:2001gc,Kosowsky:2001xp}, while the infrared cutoff $k_{\rm IR}
$ corresponds to the longest mode that reenters the horizon at the end of the kination epoch~\cite{Bhaumik:2025kuj}.  For a transition extending from the end of the pure-kination stage at $a_{\rm kin}$ to the onset of radiation domination at $a_{\rm RD}$, the infrared scale can be written as
\begin{equation}\label{S43}
		k_{\rm IR}
		=
		a_{\rm eq}\sqrt{H_{\rm eq}H_{\rm kin}}\,
		\exp\!\left[
		\frac{1}{4}
		\int_{\ln a_{\rm kin}}^{\ln a_{\rm RD}}
		[3w(a)-1]\,d\ln a
		\right].
\end{equation} 
Here, we have used $d\ln H/d\ln a=-3[1+w(a)]/2$ and $H\propto a^{-2}$ during radiation domination. In Eq.~(\ref{S43}), subscript `eq' denotes the epoch of matter-radiation equality,     $a_{\rm RD}$ represents the scale factor at the beginning of the  radiation-dominated stage, and  $w(a)$ is  the equation of state parameter between the kination and radiation-dominated epochs. In the sharp-transition approximation adopted in the present calculation, one has $a_{\rm RD}=a_{\rm kin}$, which indicates that $k_{\rm IR} \sim a_{\rm kin}H_{\rm kin}$.

The cutoffs restrict the internal momenta as $k_{\rm IR}<q<k_{\rm UV}$ and $k_{\rm IR}<|\mathbf{k}-\mathbf{q}|<k_{\rm UV}$. We evaluate the external spectrum only for $k>k_{\rm IR}$ because the present kernel is derived for the pure-kination stage.  Under these restrictions on the internal momenta, the power spectrum of curvature perturbations derived in Appendix~\ref{AB}  
   can be written as
\begin{equation}
	\begin{aligned}
		\mathcal{P}_{\mathcal{R}}(k)
		={}&
\frac{1024\,C^4G^4k^2}{81}
		\int_{k_{\rm IR}/k}^{k_{\rm UV}/k}du
	\int_{\max\left\{ \lvert1-u\rvert,k_{\rm IR}/k\right \}}
	^{\min\left \{1+u,k_{\rm UV}/k\right \} }dv 
		\left|\Pi^{(V)}(uk)\right|^2
		\left|\Pi^{(V)}(vk)\right|^2
		\mathcal{T}(u,v).
	\end{aligned}
\end{equation}
Near the infrared cutoff, $k\sim k_{\rm IR}\ll k_{\rm UV}$,
one has $k_{\rm IR}/k\sim 1$ and $k_{\rm UV}/k\gg1$.
The integration domain of $u$ is therefore dominated by the region
$u\gtrsim1$.  Now, we assume that the integral of $u$ is dominated by  the  region away from the infrared and ultraviolet
boundaries, which indicates that the integration interval of $v$ reduces approximately to
$u-1<v<u+1$. Then  we can  obtain
\begin{equation}
	\label{C3}
	\begin{aligned}
		\mathcal{P}_{\mathcal{R}}(k)
		\simeq{}&
		\frac{1024\,C^4G^4k^2}{81}
		\int_{1}^{k_{\rm UV}/k}du
		\int_{u-1}^{u+1}dv
		\left|\Pi^{(V)}(uk)\right|^2
		\left|\Pi^{(V)}(vk)\right|^2
		\mathcal{T}(u,v).
	\end{aligned}
\end{equation}

When $u\gg1$,  we have 
$
\int_{u-1}^{u+1} dv\, f(u,v)\simeq 2 f(u,u),
$
which indicates that Eq.~(\ref{C3}) can be simplified to be 
\begin{equation} \label{C33}
	\begin{aligned}
		\mathcal P_{\mathcal R}(k)
		&\simeq
		\frac{2048\,C^4G^4k^2}{81}
		\int_{1}^{k_{\rm UV}/k} du\,
		|\Pi^{(V)}(uk)|^4\,
		\mathcal T(u,u) \\
		&\simeq
		\frac{2048\,C^4G^4k^2}{81}
		\int_{1}^{k_{\rm UV}/k} du\,
		u^6  |\Pi^{(V)}(uk)|^4  \\
		&\simeq		\frac{2048\,C^4G^4k_{\rm IR}^7}{81\,k^5}
		\int_{1}^{k_{\rm UV}/k_{\rm IR}} dy\,
		y^6 \left[|\Pi^{(V)}(yk_{\rm IR})|^2\right]^2, 
	\end{aligned}
\end{equation}
where $y\equiv \frac{uk}{k_{\rm IR}}$. Eq.~(\ref{C33})  shows clearly that 
\begin{equation}
	\mathcal P_{\mathcal R}(k)\propto k^{-5}.
\end{equation}
Thus, the spectral index of the power spectrum of curvature perturbations takes the form 
\begin{equation}
	n\equiv \frac{d\ln \mathcal{P}_{\mathcal R}}{d\ln k} \simeq -5. 
\end{equation}


\section{Primordial Magnetic Fields}
\label{AE}

In Fourier space, the general ESET can be decomposed into~\cite{Mack:2001gc}
\begin{equation}
	\begin{aligned}
		\Pi^{(B)}_{ij}(\mathbf{k})
		=\frac{1}{3}\delta_{ij}{\Pi}
		+\left(\frac{k_i k_j}{k^2}-\frac{1}{3}\delta_{ij}\right)\Pi^{(S)}(\mathbf{k}) 
		+\frac{k_i}{k}\Pi^{(V)}_j(\mathbf{k})
		+\frac{k_j}{k}\Pi^{(V)}_i(\mathbf{k})
		+\Pi^{(T)}_{ij}(\mathbf{k}) ,
	\end{aligned}
\end{equation}
where $\Pi$ and $\Pi^{S}$ are the isotropic and anisotropic scalar  parts, respectively, and  $\Pi^{(V)}$ and $\Pi^{(T)}$ are respectively  the vector and tensor components, which  satisfy $k^i\Pi^{(V)}_i=k^i\Pi^{(T)}_{ij}=\Pi^{(T)i}_{\ \ i}=0$.

The anisotropic scalar, vector and tensor  components can be obtained through the following projections
\begin{equation}
	\Pi^{(S)}(\mathbf{k})
	=
	\frac{3}{2}
	\left(\frac{k_i k_j}{k^2}-\frac{1}{3}\delta_{ij}\right)
	\Pi^{(B)}_{ij}(\mathbf{k}), 
\end{equation}
\begin{equation}
	\Pi^{(V)}_i(\mathbf{k})
	=
	P_{in}(\mathbf{k})\,\frac{k_m}{k}\,\Pi^{(B)}_{mn}(\mathbf{k}),
\end{equation}
and 
\begin{equation}
	\Pi^{(T)}_{ij}(\mathbf{k})
	=
	\Lambda_{ijmn}(\mathbf{k})\Pi^{(B)}_{mn}(\mathbf{k}),
\end{equation}
where 
\begin{equation}
	P_{ij}(\mathbf{k})\equiv \delta_{ij}-\frac{k_i k_j}{k^2},
\end{equation}
and 
\begin{equation}
	\Lambda_{ijmn}(\mathbf{k})
	=
	\frac{1}{2}
	\left[
	P_{im}(\mathbf{k})P_{jn}(\mathbf{k})
	+
	P_{in}(\mathbf{k})P_{jm}(\mathbf{k})
	\right]
	-\frac{1}{2}
	P_{ij}(\mathbf{k})P_{mn}(\mathbf{k}) .
\end{equation}
It is easy to see that the scalar, vector, and tensor parts are different irreducible
projections of the same ESET and they can source different
metric sectors at linear order.  

If  the  ESET arises from the PMFs, it can be expressed as 
\begin{equation}\label{E7}
	\begin{aligned}
		\Pi^{(B)}_{mn}(\mathbf{k})
		=&\frac{1}{32\pi^4}\int d^3q \Bigl[
		B_m(\mathbf{q})B_n(\mathbf{k}-\mathbf{q})
		-\frac{1}{2}\delta_{mn}B_l(\mathbf{q})B_l(\mathbf{k}-\mathbf{q})
		\Bigr], 
	\end{aligned}
\end{equation}
which shows explicitly that the magnetic stress-energy tensor is quadratic of the magnetic field.   

For a stochastic PMF, its characteristics can be described by the power spectrum $P_B$, which is defined as 
\begin{equation}\label{E8}
	\left\langle B_i(\mathbf{k})B_j(\mathbf{k}')\right\rangle
	=(2\pi)^3P_{ij}(\mathbf{k})P_B(k)\delta(\mathbf{k}+\mathbf{k}').
\end{equation}
Since a power-law form of power spectrum offers a minimal parametrization of the magnetic power distributed over a continuous range of comoving scales, we parameterize the magnetic spectrum as a finite-band power-law spectrum,
\begin{equation} \label{E9}
	P_B(k)=A_B k^{n_B}\Theta(k_{\rm UV}-k)\Theta(k-k_{\rm IR}),
\end{equation}
where $k_{\rm UV}$ and $k_{\rm IR}$ denote the largest and smallest
comoving wavenumbers over which the magnetic field is supported.  This
parametrization has been widely used in studies of PMF-induced cosmological
perturbations~\cite{Kosowsky:2001xp,Mack:2001gc,Paoletti:2008ck,Planck:2015zrl}, and is also naturally connected with inflationary magnetogenesis
scenarios~\cite{Ratra:1991bn,Martin:2007ue,Subramanian:2015lua,Yokoyama:2015era}. For a magnetic field from the inflationary magnetogenesis, the ultraviolet cutoff can be set by the comoving
Hubble scale at the end of inflation,
$k_{\rm UV}\sim a_{\rm inf}H_{\rm inf}$.  The infrared cutoff
$k_{\rm IR}\sim H_{\rm kin}a_{\rm kin}$ corresponds to the longest mode that
reenters the horizon during the kination era~\cite{Bhaumik:2025kuj}.

\endgroup

\bibliography{ref_dedup}


\end{document}